\documentstyle[12pt]{article}
 \title{
Some theorems in thermoelasticity 
for micropolar porous media 
} 
 \author{
 \it Gerardo Iovane \qquad \qquad Francesca Passarella
} 
 \date{ 
 \small
 Dipartimento di Ingegneria dell'Informazione 
 e Matematica Applicata \\
 \small
 Universit\'a di Salerno - via Ponte don
 Melillo - I--84084 Fisciano (SA)
}

 \rightmargin=\leftmargin

 \oddsidemargin - 0.5 pt 
 \evensidemargin - 0.5 pt 

 \chardef \i = "10

 \def\c#1{\setbox0=\hbox{#1}\ifdim\ht0=1ex \accent'30 #1%
\else{\ooalign{\hidewidth\char'30\hidewidth\crcr\unhbox0}}\fi}

\begin {document}
 \maketitle
 
 \begin{abstract}
Within the context of a linear theory of heat--flux dependent thermoelasticity  
for 
micropolar porous media some variational principles and a reciprocal
relation  are derived.
 \end{abstract}
 
 \section{Introduction}

This paper is concerned with the linear theory of the generalized thermodynamics
for the micropolar porous media.

 The {\it micropolar elasticity} theory has been established
by Eringen [1-4] and is of fundamental interest in a lot of cases for which the
classical theory of elasticity is inadequate. Other related results have been obtained in
[5-7].

A general theory of  materials with voids has been developed by Cowin and Nunziato in
[8-9], by using the concept of {\it distributed body} introduced in
 \cite{[10]}.
In this theory, the matrix material is elastic and the
interstices are void of material.
 In this context, the {\it bulk density} 
is written as the product of two fields: the material density field and
the volume fraction field. 
The intended applications of this theory  are to
geological materials (like rock and soils) and to 
manufactured porous materials
(like ceramics and pressed powders).

In a previous paper \cite{[11]}, we studied the thermodynamic theory of
micropolar porous materials by using the theories developed in [1-4],
\cite{[6]} and in [8-9]. This study was based on a generalized theory
of thermoelasticity including the {\it heat flux vector} in the set of
the constitutive variables. In \cite{[11]}, we also derived  the
corresponding linear theory.

In the present work, starting from the basic equations 
established in \cite{[11]},
 we state two variational characterization 
(of Biot and Hamilton types, see [12], [13]) of the initial--boundary value
problem.
In final part, we also
derive a reciprocal relation for the problem in concern
 by using convolution in
time (fundamental results, in this context, have been  obtained by  D. Graffi
 \cite{[15]})  and Laplace trasform.

 \section {Basic Equations}

Throughout this paper, by $ \Omega $ and 
 $ \partial \Omega $ we shall denote the region and its boundary of the
physical $ 3 $--dimensional space $ ( \equiv \Re^3) $ 
occupied by an elastic solid in a given 
re\-ference (unstressed) state and $I=[0,\,+\infty)$. We refer the motion of
the body  to a fixed orthonormal frame in $ \Re^3 $.
We shall denote the tensor components of order $ p \geq 1 $ by Latin 
subscripts,
ranging over $ \{1, 2, 3 \} $. Summation over repeated subscripts is implied. 
Superposed dots or subscripts preceded by a comma will mean partial
derivative with respect to the time or the corresponding coordinates.

On the basis of the linear theory established in
\cite{[11]},  the behaviour of a micropolar porous thermoelastic
body is governed by the following local balances of momentum,
equilibrated force and energy
 \newline
 \begin{equation}
 \label{1}
 \begin{array}{l}
 T_{ji, j} + \rho f_{i} = \rho \ddot{u}_{i},
\\[7 mm]
 M_{ji, j} + \epsilon_{irs}T_{rs} + \rho  \ell_i ,
 = \rho J_{ij} \ddot{ \phi}_{j}
\end{array}
 \end{equation}
$$
 \begin{array}{l}
h_{i, i} + g + \rho \ell = \rho \stackrel{ \,}{ \chi} \ddot{ \varphi},
\\[7 mm]
 \rho( \theta_0 \dot{ \eta} - r) = q_{i,i},
\end{array}
$$
 \newline
where 
 $$ 
 \begin{array}{ll}
 T_{ij}, \, M_{ij} 
&
 \hbox{stress and couple stress tensors;} \\
h_i, \, g
&
 \hbox{equilibrated stress vector, intrinsic equilibrated body force;} 
 \\
 \eta, \, q_i 
&
 \hbox{entropy and heat flux vector.} 
 \end{array}
 \qquad \qquad \qquad \qquad \qquad \qquad 
 $$ 
These are the dependent variables of the theory, needing a constitutive 
equation.

Further,
 $ \rho $ is the bulk mass density, $ f_i $ the body force,
 $  \ell_i $ the body couple, $ J_{ij} $ the micro inertia tensor,
 $ \stackrel{ \,}{ \chi} $ the equilibrated inertia,
 $ \ell $ the extrinsic equilibrated body force,
 $ r $ the strength of the internal heat source and
 $ \theta_0 $ the (constant $ > 0$) temperature in the reference state;
 $ \epsilon_{irs} $ is the alternating symbol.

Finally, $ u_i, \, \phi_i $ and $ \varphi $, along with the temperature 
 $ \theta $ (measured from $ \theta_0 $), are the thermokinetic
 variables of the theory: $ u_i $ represents the 
displacement, $ \phi_i $ the microrotation, $ \varphi $ the change in volume
fraction from the reference configuration.
 
The constitutive equations for (\ref{1}) are \cite{[11]}
 \newline
 \begin{equation}
 \label{2}
 \begin{array}{l}
T_{ij} = C_{ijkl}E_{kl} + G_{ijkl} { \mit \Psi}_{lk} + 
 H_{ij} \varphi + 
 H_{ijk} \varphi_{,k} + 
 A_{ij} \theta,
\\[7 mm]
M_{ij} = G_{klij} E_{kl}
 + { \mit \Gamma }_{ijkl}{ \mit \Psi}_{lk} + P_{ij} \varphi + 
P_{ijk} \varphi_{,k} + G_{ij} \theta ,
\\[7 mm]
g = - H_{ij}E_{ij} - \,P_{ij}{ \mit \Psi}_{ji} - a \varphi 
 - a_i \varphi_{,i} - b \theta,
\\[7 mm]
h_i = H_{jki}E_{jk} + P_{jki}{ \mit \Psi}_{kj} + 
a_i \varphi + D_{ij} \varphi_{,j} + \gamma_i \theta,
\\[7 mm]
 \rho \eta = - A_{ij}E_{ij} - \,G_{ij}{ \mit \Psi}_{ji}
 - b \varphi - 
 \, \gamma_i \varphi_{,i} + c \theta,
\\[7 mm]
(1 + \tau \frac{ \partial \;}{ \partial t})q_i \, = \, K_{ij} \theta_{,j}.
 \end{array}
 \end{equation}
 \newline
Here, $ E_{ij} $ and $ { \mit \Psi}_{ij} $ are the {\it deformation} and 
{\it wryness tensors} defined by
 \newline
 \begin{equation}
 \label{3}
E_{ij} = u_{j,i} - \epsilon_{ijh} \phi_h, \qquad \qquad \qquad
{ \mit \Psi}_{ij} = \phi_{i,j}.
 \end{equation} 
 \newline
We shall consider a homogeneous body; in this case, the coefficients in
(\ref{2}) are all constants; it can also be proved 
 \cite{[11]} that the following symmetry relations follow
 \newline
 \begin{equation}
 \label{4}
 \begin{array}{l}
C_{ijkl} = C_{klij} ,
 \qquad { \mit \Gamma }_{ijkl} = { \mit \Gamma }_{klij} 
 \qquad
D_{ij} = D_{ji}
 \end{array}
 \end{equation}
 \newline
and that the conductivity tensor $ K_{ij} $ is 
inversable.

Note that applying the operator 
 $ 
 \left(1 + \tau \frac{ \partial}{ \partial t} \right)
 $ 
to the energy equation, by
(\ref{2}$ {}_{5,6} $) we get an evolution equation for $ \theta $ 
of hyperbolic type
\newline
\begin{equation}
 \label{5}
K_{ij} \theta_{,ji} + \left( 1 + \tau \frac{ \partial}{ \partial t}
 \right)
 [ \rho r - \rho \theta_0 \dot{ \eta} ] = 0.
 \end{equation} 
 \newline
By insertion of (\ref{2}, \ref{3})
 into (\ref{1}${}_{1,2,3}$, \ref{5}), we obtain the complete system of
(coupled) field equations for $ u_i $, $ \phi_i $, $ \varphi $ and $
\theta $.

Let us  denote four pairs of disjoint and complementary subsets of the
(smooth) boundary $ \partial \Omega $ by
 $ \{ \Sigma_i , \Sigma_{i + 1} \} $ 
with $ i = 1, 3, 5, 7 $ and denote the outward unit normal
to $ \partial \Omega $ by $ n_j $.

To the system of field equations (\ref{1}${}_{1,2,3} $ , \ref{2} ,
 \ref{5}) we add the following initial--boundary conditions
 \newline
 \begin{equation}
 \label{6} 
 u_i = 
 \dot{u}_i = 
 \phi_i = 
 \dot{ \phi_i} = 
 \varphi = 
 \dot{ \varphi} = 
 \theta = 
 q_i = 0 
 \qquad \quad \hbox{in} \; \Omega \times \{0 \},
 \end{equation}
\newline
$$
 \begin{array}{lrlr}
u_i = u_i^{*} 
& \; \, \hbox{on} \; \Sigma_1 \times  I  ,
& \qquad \qquad
T_{ji}n_j = t_i^{*}
& \; \, \hbox{on} \; \Sigma_2 \times  I  ,
 \end{array}
$$
\begin{equation}
 \label{7} 	
 \begin{array}{lrlr}
 \phi_i = \phi_i^{*} 
& \; \, \hbox{on} \; \Sigma_3 \times  I  ,
& \qquad \qquad
 M_{ji}n_j = m_i^{*}
& \; \, \hbox{on} \; \Sigma_4 \times  I  ,
\\[7 mm]
\varphi = \varphi^{*}
& \; \, \hbox{on} \; \Sigma_5 \times  I  ,
& \qquad \qquad
 h_j n_j = h^{*}
& \; \, \hbox{on} \; \Sigma_6 \times  I  ,
\\[7 mm]
 \theta = \theta^{*}
& \; \, \hbox{on} \; \Sigma_7 \times  I  ,
& \qquad \qquad
 q_j n_j = q^{*}
& \; \, \hbox{on} \; \Sigma_8 \times  I  , 
 \end{array}
 \end{equation}
 \newline
where
right-hand terms stand for (sufficiently smooth) assigned 
fields; along with $ f_i $, $  \ell_i $, $ \ell $ and $ r $, these are the {\it data} of
the mixed
 problem considered.
 An array field $ \bigl( u_i, \; \phi_i, \; \varphi , \;
\theta \bigr) $ 
 meeting all equations
 (\ref{1}${}_{1,2,3} $, \ref{2},
 \ref{5}, \ref{6}, \ref{7}), for some assignment of the 
data, will be referred to as a (regular) {\it solution} of this.

The uniqueness of solution to the above problem has
been established in
 \cite{[11]}.


 \section{Variational principle I}
In this section, following \cite{[13]}, we establish a variational theorem of 
 Hamilton type.
We introduce (see also \cite{[11]}) the {\it kinetic  energy}  $ K $ and
the {\it free energy} $ \psi $ 
  \newline
\begin{equation}
 \label{8}
 \begin{array}{rl}
 \rho K = & \, \, 
 \frac 12 \rho ( \dot{u}_i \dot{u}_i
 + J_{ij} \dot{ \phi}_i \dot{ \phi}_j
 + \stackrel{ \,}{ \chi} \dot{ \varphi}^2)
\\[7 mm]
 \rho \psi = & \, \,
 \frac 12 C_{ijkl} E_{ij} E_{kl} + 
 \frac 12 { \mit \Gamma }_{ijkl} { \mit \Psi}_{ji}{ \mit \Psi}_{lk}
 + \frac 12 a \varphi^2 
 + \frac 12 D_{ij} \varphi_{,i} \varphi_{,j}
 - \, \frac 12 c \theta^2 + 
\\[7 mm]
& + 
 \frac 12 B_{ij} q_i q_j + 
 G_{ijkl} E_{ij} { \mit \Psi}_{lk} + 
 H_{ij} E_{ij} \varphi + 
 H_{ijk} E_{ij} \varphi_{,k} + 
 A_{ij} E_{ij} \theta + 
 \\[5  mm]
& + 
 P_{ij} { \mit \Psi}_{ji} \varphi + 
P_{ijk}{ \mit \Psi}_{ji} \varphi_{,k} 
 + G_{ij}{ \mit \Psi}_{ji} \theta
 + 
a_i \varphi \varphi_{,i} + 
b \varphi \theta + 
 \gamma_i \varphi_{,i} \theta.
 \end{array}
 \end{equation}
 \newline
We can prove the following 

 \subparagraph{Variational principle I}
{\it Let 
$ t_1 $ and $ t_2 $ be arbitrary instants of time in $I$ and
assume that

 i. \quad the symmetry relations (\ref{4}) hold;
 
 ii. \quad $ K_{ij} $ and $ J_{ij} $ are symmetric tensors.
 \par

Then, for arbitrary variations in $ u_i $, $ \phi_i $, $ \varphi $,
 $ \theta $ such that 
}
\newline
 \begin{equation}
 \label{9}
 \delta \, u_i = 0 , \quad \,
 \delta \, \phi_i = 0 , \quad \,
 \delta \, \varphi = 0 , \quad \,
 \delta \, \theta = 0 , \quad \,
 \hbox{in} \; \Omega \times \{t_1, \,t_2\},
 \end{equation}
\smallskip
\begin{equation}
 \label{10}
 \begin{array}{lrlr}
 \delta \,u_i = 0 
& \; \, \hbox{on} \; \Sigma_1 \times  I  ,
& \qquad \qquad
 \delta \, \phi_i = 0
& \; \, \hbox{on} \; \Sigma_3 \times  I  ,
\\[7 mm]
 \delta \, \varphi = 0
& \; \, \hbox{on} \; \Sigma_5 \times  I  ,
& \qquad \qquad
 \delta \, \theta = 0
& \; \, \hbox{on} \; \Sigma_7 \times  I , 
 \end{array}
 \end{equation}
\newline
{\it and the functions $ f_i, \,  \ell_i, \, \ell $, $ \eta $, $ r $, 
 $ t_i^{*} $, $ m_i ^{*} $ ,
 $ h^{*} $, $ q^{*} $, $t_1$ and $t_2$ being kept unchanged,
the following variational equations
}
 \newline
 \begin{equation}
 \label{11}
 \begin{array}{ll}
 \delta 
 \displaystyle
 \int_{t_1}^{t_2}
\,
dt
& \! \! \! \! \! \!
 \left[
\displaystyle \int_{ \Omega }
 \rho \Bigl(
K - \psi - \eta T + \frac 1{2\rho} B_{ij}q_iq_j
 + 
 f_iu_i
 + 
 \ell_i \phi_i
 + 
 \ell  \varphi
 \Bigr) \, dV 
 \right.
 + 
\\[7 mm]
&
 \! \! \!
 + 
 \left.
 \displaystyle
 \int_{ \Sigma_2}
t_i^{*}u_i 
 \, d \sigma 
 + 
 \int_{ \Sigma_4}
m_i^{*} \phi_i
 \, d \sigma 
 + 
 \int_{ \Sigma_6}
h^{*} \varphi 
 \, d \sigma 
 \right] \, \, = \, 0
 \end{array}
 \end{equation}
 \newline
 \begin{equation}
 \label{12}
 \delta \int_{t_1}^{t_2}dt
 \left[
 \int_{ \Omega } 
 \frac 12 K_{ij}T_{,i} T_{,j} \, dV - (1
 + \tau p) \left \{
 \int_{ \Omega }
 \rho ( \eta \dot{T} - r)T \, dV + 
 \int_{ \Sigma_8 }
q^* T \, d \sigma 
 \right \} \right] = 0
 \end{equation}
 \newline
{\it
are satisfied if and only if are verified the field 
equations (\ref{1}${}_{1,2,3} $) and (\ref{5}).
In the above equations we put $ T = \theta + \theta_0 $ 
(the absolute temperature) and 
 $ p \equiv \displaystyle \frac { \partial \;}{ \partial t} $
}
.

 \subparagraph{Proof.}
We can remark that, under the conditions (\ref{9}), the variation of the
kinetic energy  is
 \newline
 \begin{equation}
 \label{13}
 \delta \int_{t_1}^{t_2}
dt \int_{ \Omega }
 \rho K \, dV = - \int_{t_1}^{t_2}
dt
 \int_{ \Omega }
 \rho
 \Bigl( \ddot{u}_i \delta u_i
 + J_{ij} \ddot{ \phi}_j \delta \phi_i
 + \stackrel{ \,}{ \chi} \ddot{ \varphi} \delta \varphi
 \Bigr) \, dV.
 \end{equation}
 \newline
Moreover, using (\ref{8}${}_2 $, 
 \ref{2}, \ref{3}, \ref{4}, \ref{7}, \ref{9}, \ref{10}) and the divergence
theorem, we can obtain 
 \newline
 \begin{equation}
 \label{14}
 \begin{array}{rl}
 \delta 
 \displaystyle
 \int_{t_1}^{t_2}
& \! \! \! \!
dt
 \displaystyle
 \int_{ \Omega }
\left(
 \rho \psi + \rho \eta T - \frac 12 B_{ij}q_iq_j 
\right)
\, dV 
 = 
\\[7 mm]
& \! \! \! \!
 = \, \,
 \displaystyle
 \int_{t_1}^{t_2}
dt
 \int_{ \Omega }
 \bigl( T_{ji} \delta E_{ji} + M_{ji} \delta { \mit \Psi}_{ij} - g
 \delta \varphi + h_i \delta \varphi_{,i}) dV = 
\\[7 mm]
& \! \! \! \!
 = 
 \displaystyle
 \int_{t_1}^{t_2}dt
 \left[
 \int_{ \Omega }
 \Bigl\{
 - T_{ji,j} \delta u_i
 - \bigl(
M_{ji,j} + \epsilon_{irs}T_{rs}
 \bigr) 
\delta \phi_i
 - \bigl(
h_{i,i} + g
 \bigr) \delta \varphi
 \Bigr \} \, dV 
 \right.
 + 
\\[7 mm]
& \! \! \!
 \qquad
 + 
 \left.
 \displaystyle
 \int_{ \Sigma_2}
t_i^{*} \delta u_i 
 \, d \sigma 
 + 
 \int_{ \Sigma_4}
m_i^{*} \delta \phi_i
 \, d \sigma 
 + 
 \int_{ \Sigma_6}
h^{*} \delta \varphi 
 \, d \sigma 
 \right].
 \end{array}
 \end{equation}
Taking into account  (\ref{13}) and (\ref{14}), the variational 
equation (\ref{11}) can be rewritten as follows
 \newline 
$$ 
 \begin{array}{rl}
 \displaystyle \int_{t_1}^{t_2}
dt
 \int_{ \Omega }
 \Bigl[
 \bigl(T_{ji,j} + \rho f_i - \rho \ddot{u}_i \bigr) \delta u_i
 + \bigl(
M_{ji,j} + \epsilon_{irs}T_{rs} + \rho \ell_i
 - \rho J_{ij} \ddot{ \phi}_j
 \bigr)
 \delta \phi_i
&
\\[7 mm]
 + \bigl(
h_{i,i} + g + \rho \ell - \rho  \stackrel{ \,}{ \chi}\ddot{ \varphi}
 \bigr) \delta \varphi \Bigr] \, dV 
& = \,0
 \end{array}
 $$ 
\newline 
This equation holds if and only if (\ref{1}${}_{1,2,3} $) are satisfied.

From  (\ref{2}${}_6 $) and  (\ref{10}) it follows that
\newline 
 $$ 
 \delta 
 \displaystyle 
 \int_{t_1}^{t_2}
dt
 \int_{ \Omega } 
 \frac 12 K_{ij}T_{,i} T_{,j} \, 
dV = 
 \int_{t_1}^{t_2}
dt
 \left[
(1 + \tau p)
 \int_{ \Sigma_8}
q^* \delta \theta \, d \sigma 
 - 
 \int_{ \Omega } 
 K_{ij} \theta_{,ij} \delta \theta \, dV 
 \right]
 $$ 
\newline 
where $p$ has been treated as a constant \cite{[12]}.

Thus,  (\ref{12}) can be written as
\newline 
 $$ 
 \int_{t_1}^{t_2}
dt
 \int_{ \Omega } 
 \left \{
K_{ij} \theta_{,ij} \, + \rho
(1 + \tau p)
 (r - T \dot{ \eta})
 \right \} \delta \theta \, dV 
 = 
 - 
 \int_{ \Omega } 
 \left. ( \eta T \delta T)
 \right|_{t_1}^{t_2} \, dV
 $$ 
\newline 
After the linearization, from (\ref{9}) we conclude that
 \newline 
$$ 
 \int_{t_1}^{t_2}
dt
 \int_{ \Omega } 
 \left \{
K_{ij} \theta_{,ij} \, + \rho
(1 + \tau p)
 (r - \theta_0 \dot{ \eta}) \right \} \delta \theta \, dV 
 \, = 
 \,0
 $$
 \newline
This equation holds if and only if (\ref{5}) is satisfied.

\section{Variational pinciple II}

Following \cite{[12]},
we introduce the {\it entropy flow} vector $ s_i $  by
 $$ 
 \theta_0 \dot{s}_{i} = q_i
 $$ 
and by (\ref{1}${}_4 $, \ref{2}${}_5 $) we get 
 \newline
 \begin{equation}
 \label{15}
 \dot{s}_{i,i} = 
 - A_{ij} \dot{E}_{ij} - G_{ij} \dot{ \mit \Psi}_{ji} - b \dot{ \varphi}
 - \gamma_i \dot{ \varphi}_{,i} + c \dot{ \theta}
 - \frac{ \rho}{ \theta_0} r.
 \end{equation}
 \newline
In the  following, we shall consider  prescribed $ f_i $, $ \ell_i $, $
\ell $ and $ r $ in $ \Omega $, and 
 $ u_i $ on $ \Sigma_1 $, $ T_{ij} $ on $ \Sigma_2 $, 
 $ \phi_i $ on $ \Sigma_3 $, $ M_{ij} $ on $ \Sigma_4 $, 
 $ \varphi $ on $ \Sigma_5 $, $ h_i $ on $ \Sigma_6 $, $ \theta $ on $ \Sigma_7 $ 
and $ q_i $ on $ \Sigma_8 $. Moreover, 
we also consider $s_i$ assigned on $\Sigma_8$.

If we assume that $ K_{ij} $ is symmetric and denote its inverse by
 $ \widetilde{K}_{ij}  \Bigl( = \widetilde{K}_{ji} \Bigr) $, we set
\newline 
 $$ 
D \, = \, \frac 12 \displaystyle \int_{ \Omega } \theta_0 (p + \tau p^2) 
 \widetilde{K}_{ij} s_i s_j \, dV 
 $$
 \newline 
so the variation of this functional due to variation of $ s_i $ is
 \newline
 \begin{equation}
 \label{16}
 \delta D \, = \, \int_{ \Omega } \theta_0 (p + \tau p^2) 
 \widetilde{K}_{ij} s_i \delta s_j \, dV .
 \end{equation}
 \newline
Here, $p$ is used as in section $3$.

We now introduce the following functional
\newline 
$$ 
V = \, \int_{ \Omega }
\bigl( \rho \psi + \rho \eta \theta -
 \frac 12 
B_{ij}q_i q_j
\bigr) \, dV .
$$ 
\newline 
If we consider  variations of $ V $ due to
variations of $ u_i $, $ \phi_i $, 
$ \varphi $, $ \theta $ and use
(\ref{8}${}_{2}$),  the constitutive equations and the
symmetry relations (\ref{4}), we obtain 
 \newline
 \begin{equation}
 \label{17}
 \begin{array}{rl}
 \delta V \, = \,
 \displaystyle \int_{ \Omega } &
\! \! \! \! \!
 \Bigl \{
c \theta \delta \theta + 
 \bigl(T_{ij} - A_{ij} \theta \bigr) \delta E_{ij} + 
 \bigl(M_{ij} - G_{ij} \theta \bigr) \delta { \mit \Psi}_{ji} + 
\\[7 mm]
&
\! \! \! \! \!
 - \bigl( g + b \theta \bigr) \delta \varphi 
 + \bigl(h_i - \gamma_{i} \theta \bigr) \delta \varphi_{,i}
 \Bigr \} 
 \, dV .
 \end{array}
 \end{equation}
 \newline
Moreover, we define the following functional
\newline 
$$ 
G \, = - \displaystyle
 \int_{ \Sigma_2} t_i^{*} u_i \, d \sigma 
 - \int_{ \Sigma_4} m_i^{*} \phi_i \, d \sigma 
 - \int_{ \Sigma_6} h^{*} \varphi \, d \sigma 
 - \int_{ \Sigma_7} \theta^{*} s_i n_i \, d \sigma 
 $$ 
\newline 
If  $ \delta u_i $, $ \delta \phi_i $ ,
 $ \delta \varphi $, $ \delta s_i $ vanish on $ \Sigma_1 $, $ \Sigma_3 $, 
 $ \Sigma_5 $, $ \Sigma_8 $, respectively, by  (\ref{7}) we can write 
 \newline 
$$ \delta G = 
 - \int_{ \partial \Omega } 
 \Bigl (T_{ji} \delta u_{i} + 
M_{ji} \delta \phi_{i} + 
h_j  \delta \varphi + \theta \delta s_j
 \Bigr)  n_j \, d \sigma ; 
 $$ 
\newline 
and, from the divergence theorem and (\ref{3}, \ref{4}, \ref{15}), 
we get
 \newline
\begin{equation}
 \label{18}
 \begin{array}{rl}
 \delta G = \, - 
 \displaystyle
 \int_{ \Omega }
 \Bigl \{
 &
 \! \! \! \! \!
 (T_{ij} - A_{ij} \theta) \delta E_{ij} 
 + (M_{ij} - G_{ij} \theta) \delta { \mit \Psi}_{ji} 
 + (h_i - \gamma_i \theta) \delta \varphi_{,i} 
 + c \theta \delta \theta + 
\\[7 mm]
&
 \! \! \! \! \!
 - b \theta \delta \varphi
 + T_{ji,j} \delta u_{i} 
 + \bigl(M_{ji,j} + \epsilon_{irs}T_{rs} \bigr) \delta \phi_{i} 
 + h_{j,j} \delta \varphi 
 + \theta_{,j} \delta s_j \Bigr \} \, dV .
 \end{array}
 \end{equation}

Finally, 
we put
\newline 
 $$ 
F \, = \,{ \displaystyle \int_{ \Omega }} \rho 
 \left \{ \left( \frac 12 p^2 u_i - 
f_i \right) u_i + 
 \left( \frac 12 p^2 J_{ij} \phi_j - \ell_i \right) \phi_i + 
 \left( \frac 12 \stackrel{ \,}{ \chi} p^2 \varphi - 
 \ell \right) \varphi \right \} \, dV 
,
 $$ 
\newline 
so, assuming the tensor $ J_{ij} $ symmetric, the variation of this
functional is 
 \newline
 \begin{equation}
 \label{19}
 \delta F \, = \, \int_{ \Omega } \rho
 \Bigl \{ \bigl( p^2 u_i - f_i \bigr) \delta u_i + 
 \bigl( p^2 J_{ij} \phi_j - \ell_i \bigr) \delta \phi_i + 
 \bigl( \stackrel{ \,}{ \chi} p^2 \varphi - \ell \bigr)
 \delta \varphi \Bigr \} \, dV .
 \end{equation}
 \newline
Thus, we can obtain the following variational theorem (of the Biot type)
 \subparagraph{Variational pinciple II}
 {\it 
Let us  prescribe $ f_i, \, \ell_i, \, \ell $ and $ r $ in $ \Omega $, and 
 $T_{ij}, \,M_{ij}, \,h_{i}, \, $ and $ \theta $ on $ \Sigma_2, \,
 \Sigma_4, \, \Sigma_6, \, \Sigma_7 $, respectively,
and
 assume that

 i. \quad the symmetry relations (\ref{4}) hold;
 
 ii. \quad $ K_{ij} $ and $ J_{ij} $ are symmetric tensors.
 \par

Then,
for every} $ \delta u_i $  $ ( = 0 $ on $ \Sigma_1) $, 
 $ \delta \phi_i $ $ ( = 0 $ on $ \Sigma_3) $, 
 $ \delta \varphi $ $ ( = 0 $ on $ \Sigma_5) $, 
 $ \delta s_i $ $ ( = 0 $ on $ \Sigma_8) $, 
{\it the following equation}
 \begin{displaymath}
 \delta H(u_i, \, \phi_i, \, \varphi, \, \theta) = 
 \delta V + \delta G + \delta F + \delta D \, = \,0
 \end{displaymath}
{\it 
yields the field equations} (\ref{1}${}_{1,2,3} $ ,
 \ref{5}).
 \smallskip

 \subparagraph{Proof.}
By (\ref{16}, \ref{17}, \ref{18}, \ref{19}) we can write
\newline 
 $$ 
 \begin{array}{rl}
 \delta H(u_i, \, \phi_i, \, \varphi, \, \theta) 
 = &
 \displaystyle
 \, \int_{ \Omega }
 \Bigl \{( \rho \ddot{u}_i - \rho f_i - T_{ji,j}) \delta u_i + 
\\[7 mm]
&
 + 
 \displaystyle
( \rho J_{ij} \ddot{ \phi}_i - \rho \ell_i - M_{ji,j}
 - \epsilon_{irs}T_{rs}) \delta \phi_i + 
\\[7 mm]
&
 + 
( \, \rho \stackrel{ \,}{ \chi} \ddot{ \varphi} - 
 \rho \ell  - g - h_{i,i}) \delta \varphi + 
\\[7 mm]
&
 + 
[ \theta_0 (1 + \tau p) 
 \widetilde{K}_{ij} \dot{s}_i - \theta_{,j}] \delta s_i
 \Bigr \} \, dV.
 \end{array}
 $$ 
\newline 
Vanishing of this expression for arbitrary $ \delta u_i,
 \; \delta \phi_i, \; \delta \varphi, \; \delta s_i $ 
of course implies the field  equations 
 \ref{1}${}_{1,2,3} $; further from the last term in the integral above, we get 
 \newline 
$$ 
 \theta_0 (1 + \tau p) 
 \widetilde{K}_{ij} \dot{s}_i \, = \, \theta_{,j}.
 $$ 
\newline 
If we multiply the above equation  by $ K_{hi} $ and  derive  
with respect to the $i$--th coordinate, we can  easily
deduce the field equation (\ref{5}) from (\ref{15}). This completes the proof.

 \section {Reciprocity}
Let us denote by $ a* b $ the time convolution of the
scalar fields on $ \Omega \times  I $ 
 $$ 
 (a_1* a_2)({ \bf x}, t) =
 \int_0^ta_1({ \bf x}, t - \tau) a_2({ \bf x}, \tau) \, d \tau.
 $$ 
Let  us consider the body subjected to two different sets of data
 \newline 
$$ { \cal T}^{( \alpha)} = 
 \Bigl \{
f_i^{( \alpha)}, \ell_i^{( \alpha)}, \ell^{( \alpha)}, 
r^{( \alpha)}, 
u_i^{* \, ( \alpha)} , t_i^{* \, ( \alpha)},
 \phi_i^{* \, ( \alpha) }, m_i ^{* \, ( \alpha)}
, \varphi^{* \, ( \alpha)} ,
h^{* \, ( \alpha)},
 \theta^{* \, ( \alpha)} , q^{* \, ( \alpha)}
 \Bigr \}
 $$ 
\newline 
 and null initial conditions and let be
 \newline 
$$ 
{ \cal U}^{( \alpha)} = 
 \left( u_i^{( \alpha)} , \phi_i^{( \alpha)} ,
 \varphi^{( \alpha)} 
 , \theta^{( \alpha)} 
 \right) \qquad \qquad \qquad \alpha = 1, 2,
 $$ 
\newline
the corresponding solutions. 
Moreover, we define $ T_{ij}^{( \alpha)}, M_{ij}^{( \alpha)}, h_i ^{( \alpha)}, 
 g^{( \alpha)}, \eta^{( \alpha)}, q_i^{( \alpha)} $ by means of
 (\ref{2}) for each $ \alpha = 1, 2 $.

 We prove the following theorem
 
 \subparagraph{Reciprocal Theorem} 
 {\it 
Let  $ { \cal U}^{( \alpha)} $ be solutions corresponding to different sets of data 
 $ { \cal T}^{( \alpha)} $ ( $ \alpha = 1, 2 $), and
 assume that

 i. \quad the symmetry relations (\ref{4}) hold;
 
 ii. \quad $ K_{ij} $ and $ J_{ij} $ are symmetric tensors.
 \par

Then, the following relation holds
}
 \newline
 \begin{equation}
 \label{20}
 {\cal I}_{ \alpha \beta} = 
{\cal I}_{ \beta \alpha} 
 \end{equation}
{\it where} 
\newline
 $$ 
 \begin{array}{ll}
{\cal I}_{ \alpha \beta} = 
 &
 \displaystyle
 \int_{ \Omega }
 \rho \Bigl[
 f_i^{( \alpha)}* \dot{u}_i^{( \beta)}
 + 
\ell_i^{( \alpha)}* \dot{ \phi}_i^{( \beta)}
 + 
\\[7 mm]
& \quad 
 + 
 \ell^{( \alpha)}* \dot{ \varphi}^{( \beta)}
 - 
 \frac 1{ \theta_0} r^{( \alpha)}* \theta^{( \beta)} \Bigr]
 \, dV 
 + 
\\[7 mm]
&
 - \,
 \displaystyle
 \int_{ \Sigma_1}
\dot{u}_i^{* \, ( \alpha)}* T_{ji}^{( \beta)} n_j
 \, d \sigma 
 + 
 \int_{ \Sigma_2}
t_i^{* \, ( \alpha)}* \dot{u}_i^{( \beta)} 
 \, d \sigma 
 + 
\end{array}
$$
$$ 
 \begin{array}{ll}
&
 - \,
 \displaystyle
 \int_{ \Sigma_3}
 \dot{\phi}_i^{* \,( \alpha)}* M_{ji}^{( \beta)} n_j
 \, d \sigma 
 + 
 \int_{ \Sigma_4}
m_i^{* \, ( \alpha)}* \dot{ \phi}_i^{( \beta)} 
 \, d \sigma 
 + 
\\[7 mm]
&
 - \,
 \displaystyle
 \int_{ \Sigma_5}
 \dot{\varphi}^{* \, ( \alpha)}* h_{j}^{( \beta)} n_j
 \, d \sigma 
 + 
 \int_{ \Sigma_6}
h^{* \, ( \alpha)}* \dot{ \varphi}^{( \beta)} 
 \, d \sigma 
 + 
\\[7 mm]
&
 + 
 \displaystyle
 \frac 1{ \theta_0}
 \int_{ \Sigma_7}
 \theta^{* \, ( \alpha)}* q_{j}^{( \beta)} n_j
 \, d \sigma 
 - 
 \frac 1{ \theta_0} \int_{ \Sigma_8}
q^{* \, ( \alpha)}* \theta^{( \beta)} 
 \, d \sigma 
 \end{array}
 $$ 
 \newline
for $ \alpha, \beta = 1, 2 $.

 \subparagraph{Proof.} We put
 $$ 
 \begin{array}{ll}
{ \cal F}_{ \alpha \beta} = 
 &
 \Bigl(
 \widetilde{T}_{ji}^{( \alpha)} \widetilde{u}_i^{( \beta)} + 
 \widetilde{M}_{ji}^{( \alpha)} \widetilde{ \phi}_i^{( \beta)} + 
 \widetilde{h}_j^{( \alpha)} \widetilde{ \varphi}^{( \beta)} 
 \Bigr)_{,j} + 
\\[7 mm]
&
 + \rho \bigl( \widetilde{f}_i^{( \alpha)} \widetilde{u}_i^{( \beta)} 
 + \widetilde{l}_i^{( \alpha)} \widetilde{ \phi}_i^{( \beta)} 
 + \widetilde{ \ell}^{( \alpha)} \widetilde{ \varphi}^{( \beta)} \, \bigr),
 \end{array}
$$
$$
 \begin{array}{ll}
{ \cal H}_{ \alpha \beta} = 
 &
 \Bigl(
 \widetilde{ \theta}^{( \alpha)} \widetilde{q}_i^{( \beta)} 
 \Bigr)_{,i} - \rho \widetilde{r}^{(\alpha)} 
 \widetilde{ \theta}^{( \beta)} 
 \end{array}
 $$ 
where $ \widetilde{a} $ 
is Laplace transform with respect to $ t $ of the function $ a $: 
 $ \widetilde{a} (s)= \int_0^\infty e^{-st} a(t) dt $.
If we take Laplace transform of (\ref{1}, \ref{2}) under
the hypotheses {\it i}, {\it ii} and the initial conditions (\ref{6}), we
deduce that \newline 
$$ 
 \begin{array}{rl}
{ \cal F}_{ \alpha \beta} - { \cal F}_{ \beta \alpha} = 
&
A_{ij} \bigl(
 \widetilde{ \theta}^{( \alpha)} \widetilde{E}_{ij}^{( \beta)}
 - \widetilde{ \theta}^{( \beta)} \widetilde{E}_{ij}^{( \alpha)}
 \bigr)
 + 
G_{ij} \bigl(
 \widetilde{ \theta}^{( \alpha)} \widetilde{{ \mit \Psi}}_{ji}^{( \beta)}
 - \widetilde{ \theta}^{( \beta)} \widetilde{{ \mit \Psi}}_{ji}^{( \alpha)}
 \bigr) + 
\\[7 mm]
&
 + b \bigl(
 \widetilde{ \theta}^{( \alpha)} \widetilde{ \varphi}^{( \beta)}
 - \widetilde{ \theta}^{( \beta)} \widetilde{ \varphi}^{( \alpha)}
 \bigr)
 + 
 \gamma_i \bigl(
 \widetilde{ \theta}^{( \alpha)} \widetilde{ \varphi}_{,i}^{( \beta)}
 - \widetilde{ \theta}^{( \beta)} \widetilde{ \varphi}_{,i}^{( \alpha)}
 \bigr)
\\[7 mm]
{ \cal H}_{ \alpha \beta} - { \cal H}_{ \beta \alpha} = 
&-\theta_0 s \,\Bigl[
A_{ij} \bigl(
\widetilde{ \theta}^{( \alpha)} \widetilde{E}_{ij}^{( \beta)}
-
\widetilde{ \theta}^{( \beta)} \widetilde{E}_{ij}^{( \alpha)}
\bigr)
 + 
G_{ij} \bigl(
 \widetilde{ \theta}^{( \alpha)} \widetilde{{ \mit \Psi}}_{ji}^{( \beta)}
 - \widetilde{ \theta}^{( \beta)} \widetilde{{ \mit \Psi}}_{ji}^{( \alpha)}
 \bigr) + 
\\[7 mm]
&
 + b \bigl(
 \widetilde{ \theta}^{( \alpha)} \widetilde{ \varphi}^{( \beta)}
 - \widetilde{ \theta}^{( \beta)} \widetilde{ \varphi}^{( \alpha)}
 \bigr)
 + 
 \gamma_i \bigl(
 \widetilde{ \theta}^{( \alpha)} \widetilde{ \varphi}_{,i}^{( \beta)}
 - \widetilde{ \theta}^{( \beta)} \widetilde{ \varphi}_{,i}^{( \alpha)}
 \bigr) \Bigr],
 \end{array}
 $$ 
\newline 
so
 \begin{equation}
 \label{21}
s ({ \cal F}_{ \alpha \beta}- { \cal F}_{ \beta \alpha} ) + \frac {1}{ \theta_0}
({ \cal H}_{ \alpha \beta} - { \cal H}_{ \beta \alpha }) = 0
 \end{equation} 
 \newline 
If we integrate this equation
 over $ \Omega $ and use the divergence theorem and boundary conditions, we get
  \newline
\begin{equation}
 \label{22}
 \begin{array}{l}
 \displaystyle
 \int_ \Omega
 \rho \, s \, \Bigl[
 \,
 \widetilde{f}_i^{( \alpha)} \widetilde{u}_i^{( \beta)}
 + 
 \,
 \widetilde{l}_i^{( \alpha)} \widetilde{ \phi}_i^{( \beta)}
 + 
 \widetilde{ \ell}^{( \alpha)} \widetilde{ \varphi}^{( \beta)}
 - 
 \widetilde{f}_i^{( \beta)} \widetilde{u}_i^{( \alpha)}
 + 
\\[7 mm]
 \qquad - \,
 \widetilde{l}_i^{( \beta)} \widetilde{ \phi}_i^{( \alpha)}
 - 
 \widetilde{ \ell}^{( \beta)} \widetilde{ \varphi}^{( \alpha)}
 \Bigr] - 
 \displaystyle
 \frac { \rho}{ \theta_0}
 \Bigl[
\widetilde{r}^{( \alpha)} \widetilde{ \theta}^{( \beta)}
- \widetilde{r}^{( \beta)} \widetilde{ \theta}^{( \alpha)}
 \Bigr]
 \, dV 
 + 
\\[7 mm]
 + 
 \displaystyle
 \int_{ \Sigma_1}
s \bigl(
 \widetilde{T}_{ji}^{( \alpha)} \widetilde{u}_i^{* \, ( \beta)} 
 - \widetilde{T}_{ji}^{( \beta)} \widetilde{u}_i^{* \, ( \alpha)} 
 \bigr) n_j
 \, d \sigma 
 + 
 \int_{ \Sigma_2}
s \bigl(
 \widetilde{t}_i^{* \, ( \alpha)} \widetilde{u}_i^{( \beta)} 
 - 
 \widetilde{t}_i^{* \, ( \beta)} \widetilde{u}_i^{( \alpha)} 
 \bigr) 
 \, d \sigma 
 + 
\\[7 mm]
+ 
 \displaystyle
 \int_{ \Sigma_3}
s \bigl(
 \widetilde{M}_{ji}^{( \alpha)} \widetilde{ \phi}_i^{* \, ( \beta)} 
 - 
 \widetilde{M}_{ji}^{( \beta)} \widetilde{ \phi}_i^{* \, ( \alpha)} 
 \bigr) n_j
 \, d \sigma 
 + 
 \int_{ \Sigma_4}
s
 \bigl(
 \widetilde{m}_i^{* \, ( \alpha)} \widetilde{ \phi}_i^{( \beta)} 
 - 
 \widetilde{m}_i^{* \, ( \beta)} \widetilde{ \phi}_i^{( \alpha)} 
 \bigr)
 \, d \sigma 
 +
\\[7 mm]
+ 
 \displaystyle
 \int_{ \Sigma_5} 
s \bigl(
 \widetilde{h}_{j}^{( \alpha)} \widetilde{ \varphi}^{* \, ( \beta)} 
 - \widetilde{h}_{j}^{( \beta)} \widetilde{ \varphi}^{* \, ( \alpha)}
 \bigr) n_j
 \, d \sigma 
 + 
 \int_{ \Sigma_6}
s \bigl(
 \widetilde{h}^{* \, ( \alpha)} \widetilde{ \varphi}^{( \beta)} 
 - 
 \widetilde{h}^{* \, ( \beta)} \widetilde{ \varphi}^{( \alpha)} 
 \bigr)
 \, d \sigma 
 + 
\\[7 mm]
 + 
 \displaystyle
 \frac 1{ \theta_0} \int_{ \Sigma_7}
 \bigl(
 \widetilde{ \theta}^{* \, ( \alpha)} \widetilde{q}_{j}^{( \beta)}
 - 
 \widetilde{ \theta}^{* \, ( \beta)} \widetilde{q}_{j}^{( \alpha)}
 \bigr) n_j
 \, d \sigma 
 - 
 \frac 1{ \theta_0}
 \int_{ \Sigma_8}
 \bigl(
 \widetilde { \theta}^{( \alpha)} \widetilde{q}^{* \, ( \beta)} 
 - \widetilde { \theta}^{( \beta)} \widetilde{q}^{* \, ( \alpha)} 
 \bigr)
 \, d \sigma 
 \, \, = \,0.
\end{array}
\end{equation} 
 \newline  
Finally, we obtain the desired equation
(\ref{20}) by inverting (\ref{22}) and using the convolution 
theorem for Laplace transforms.

 \bibliographystyle{plain}

\begin{thebibliography}{99}
 
 \bibitem{[1]}
A. C. Eringen: {\it Linear Theory of Micropolar Elasticity}.
 Int. Math. Mech. { \bf 15}, 909 (1966).

 \bibitem{[2]}
A. C. Eringen: {\it Foundations of Micropolar Thermoelasticity}.
Springer--Verlag, Vienna (1970).

 \bibitem{[3]}
A. C. Eringen: {\it Theory of Micropolar Elasticity}. In 
Fracture (Edited by H. Leibowitz), Vol II, 
Academic Press, New York, 622 (1968).

 \bibitem{[4]}
A. C. Eringen and C. B. Kafadar: {\it Polar Field Theories}. In 
Continuum Physics (Edited by A. C. Eringen), Vol IV, Part I,
Academic Press, New York, 1 (1976).

 \bibitem{[5]}
 M. Ciarletta and D. Ie\c{s}an: {\it Non--Classical
 Elastic Solids}. In Pitman Research Notes in Mathematics
Series (Edited by Longman Scientific \& Technical), 293
(1993). 

 \bibitem{[6]} 
 D. S. Chandrasekharaiah: {\it Heat--flux dependent micropolar
thermoelasticity}, Int. J. Engng. Sci. { \bf 24}, 1389 (1986).



 \bibitem{[7]}
 M. Ciarletta: {\it Sui processi termoelastici per continui micropolari},
Atti Sem. Mat. Fis. Univ. Modena, { \bf 39}, 103 (1991).
 

 \bibitem{[8]}
 J. W. Nunziato and S. C. Cowin: {\it A non linear theory of elastic
 materials with voids}, Arch. Rat. Mech. Anal. { \bf 72}, 175 (1979).

 \bibitem{[9]}
 S. C. Cowin and J. W. Nunziato: {\it Linear elastic materials with voids},
J. Elasticity { \bf 13}, 125 (1983).



 \bibitem{[10]}
 M. A. Goodman and S. C. Cowin: {\it A continuum theory for granular
materials}, Arch. Rat. Mech. Anal. { \bf 44}, 249 (1972).

 \bibitem{[11]}
F. Passarella: {\it Some results in micropolar thermoelasticity},
 Mechanics Research
Communications, {\bf 23} No. 4, 349 (1996).



 \bibitem{[12]}
M.A. Biot: {\it Thermoelasticity and irreversible thermodynamics}, J. Appl.
Phys. { \bf 7}, 240 (1956)

 \bibitem{[12bis]}
H. Parkus: {\it Variational Principles in Thermo-and
Magneto-Elasticity}, Springer-Verlag, Vienna,  { \bf 11} (1972)

 \bibitem{[15]}
 D. Graffi: {\it Sui teoremi di reciprocit\'a nei fenomeni non stazionari},
 Atti Accad. Sc. Ist. Bologna, Serie 11, { \bf 10}, 33
(1963).


 \bibitem{[13]}
 D. S. Chandrasekharaiah: {\it Variational and reciprocal 
principles in micropolar thermoelasticity},
 Int. J. Engng. Sci. { \bf 25}, 55
(1987).


 
 \end{thebibliography}

 \end{document}